\documentstyle[12pt]{article}

\begin{document}
\begin{titlepage}
\begin{center}
\today       \hfill    LBL-32314 \\
{\bf Revised} \hfill    UCB-PTH-92/13 \\

\vskip .5in

{\large \bf Differential Geometry on Linear Quantum Groups}
\footnote{This work was supported in part by the Director, Office of
Energy Research, Office of High Energy and Nuclear Physics, Division of
High Energy Physics of the U.S. Department of Energy under Contract
DE-AC03-76SF00098 and in part by the National Science Foundation under
grant PHY90-21139.}

\vskip .5in

Peter Schupp, Paul Watts and Bruno Zumino \\[.5in]

{\em  Department of Physics\\
      University of California\\
      and\\
      Theoretical Physics Group\\
      Physics Division\\
      Lawrence Berkeley Laboratory\\
      1 Cyclotron Road\\
      Berkeley, California 94720}
\end{center}

\vskip .5in

\begin{abstract}
An exterior derivative, inner derivation, and Lie derivative are introduced on
the quantum group $GL_{q}(N)$.  $SL_{q}(N)$ is then obtained by constructing
matrices with determinant unity, and the induced calculus is found.

\end{abstract}

\end{titlepage}

\renewcommand{\thepage}{\roman{page}}
\setcounter{page}{2}
\mbox{ }

\vskip 1in

\begin{center}
{\bf Disclaimer}
\end{center}

\vskip .2in

\begin{scriptsize}
\begin{quotation}
This document was prepared as an account of work sponsored by the United
States Government.  Neither the United States Government nor any agency
thereof, nor The Regents of the University of California, nor any of their
employees, makes any warranty, express or implied, or assumes any legal
liability or responsibility for the accuracy, completeness, or usefulness
of any information, apparatus, product, or process disclosed, or represents
that its use would not infringe privately owned rights.  Reference herein
to any specific commercial products process, or service by its trade name,
trademark, manufacturer, or otherwise, does not necessarily constitute or
imply its endorsement, recommendation, or favoring by the United States
Government or any agency thereof, or The Regents of the University of
California.  The views and opinions of authors expressed herein do not
necessarily state or reflect those of the United States Government or any
agency thereof of The Regents of the University of California and shall
not be used for advertising or product endorsement purposes.
\end{quotation}
\end{scriptsize}

\vskip 2in

\begin{center}
\begin{small}
{\it Lawrence Berkeley Laboratory is an equal opportunity employer.}
\end{small}
\end{center}

\newpage
\renewcommand{\thepage}{\arabic{page}}
\setcounter{page}{1}

\section{Introduction}
The general theory of differential calculi on quantum groups is due to
Wo\-ro\-no\-wicz \cite{W}, and a number of interesting papers have been
written since (cf. \cite{B,J}).  In this paper, we present an explicit
formulation of the differential geometry on the quantum groups $GL_{q}(N)$ and
$SL_{q}(N)$.  We will show how a calculus incorporating a closed algebra of
derivations can be introduced on these quantum groups.  We approach the subject
from a more physics-oriented perspective, presenting commutation relations
between the various matrix elements, differential operators, forms, etc.  The
Hopf algebraic nature of the subject is deemphasized; there are occasional
references to such objects as ``antipode'', but in general the focus is on a
formulation which is suitable for computations.  (A treatment of some of the
material contained here using the more mathematical structures of
quasitriangular Hopf algebrae will be presented in a forthcoming paper
\cite{SWZ}.)

Many of the conventions and notations we use can be found in \cite{RTF}, as
well as other references herein.

\section{$GL_{q}(N)$}
\subsection{The Quantum Group $GL_{q}(N)$}

The R-matrix for $GL_{q}(N)$, which of course satisfies the Yang-Baxter
equation
\begin{equation}
R_{12}R_{13}R_{23}=R_{23}R_{13}R_{12}, \label{YBE}
\end{equation}
is given in \cite{RTF} as
\begin{equation}
R_{12}=q \sum_{i} e_{ii} \otimes e_{ii}+ \sum_{i \neq j} e_{ii} \otimes e_{jj}+
\lambda \sum_{i>j} e_{ij} \otimes e_{ji},
\end{equation}
where $i,j=1,\ldots,N$, $\lambda = q-q^{-1}$, and $e_{ij}$ is the $N \times N$
unit matrix with lone nonzero element at $(i,j)$.  This matrix satisfies the
characteristic equation
\begin{equation}
\hat{R}^{2}_{12}-\lambda \hat{R}_{12}-1=0, \label{char}
\end{equation}
where $\hat{R}_{12}$ is defined by
\begin{equation}
(\hat{R}_{12})^{ij}{}_{kl} = (P_{12}R_{12})^{ij}{}_{kl} = (R_{12})^{ji}{}_{kl},
\end{equation}
with $P_{12}$ being the permutation matrix which exchanges spaces 1 and 2,
as above.  The defining representation for the quantum group $GL_{q}(N)$ is
given by matrices $A$ satisfying
\begin{equation}
R_{12}A_{1}A_{2}=A_{2}A_{1}R_{12}. \label{RAA}
\end{equation}
The determinant of such a matrix can be introduced in the following way:  let
$\{x^{i} \}$ be the $N$ coordinates of the quantum hyperplane whose
transformation group is $GL_{q}(N)$, and let $\{dx^{i}\}$ be the associated
differentials.  The commutation relations between these quantities which are
preserved under such transformations are \cite{M1,WZ,Z1}
\begin{eqnarray}
x^{j}x^{i}&=&q^{-1}(R_{12})^{ij}{}_{kl}x^{k}x^{l},\\
x^{j}dx^{i}&=&q(R_{12})^{ij}{}_{kl}dx^{k}x^{l},\\
dx^{j}dx^{i}&=&-q(R_{12})^{ij}{}_{kl}dx^{k}dx^{l}.
\end{eqnarray}
These commutation relations allow us to define the Levi-Civita tensor as
\begin{equation}
dx^{i_{1}}dx^{i_{2}} \ldots dx^{i_{N}}=\epsilon _{q}^{i_{1} i_{2} \ldots i_{N}}
dx^{1} dx^{2} \ldots dx^{N}.
\end{equation}
This tensor satisfies the relations
\begin{eqnarray}
& {(R_{0N} \ldots R_{02}R_{01})^{i_{0}i_{1}i_{2} \ldots
i_{N}}}_{j_{0}j_{1}j_{2}
\ldots j_{N}} \epsilon _{q} ^{j_{1}j_{2} \ldots j_{N}} & = \nonumber \\
& {(R_{10}R_{20} \ldots R_{N0})^{i_{0}i_{1}i_{2} \ldots
i_{N}}}_{j_{0}j_{1}j_{2}
\ldots j_{N}} \epsilon _{q}^{j_{1}j_{2} \ldots j_{N}} & = \nonumber \\
& q \delta ^{i_{0}}_{j_{0}} \epsilon _{q}^{i_{1}i_{2} \ldots i_{N}}. &
\end{eqnarray}
The quantum determinant of $A$, $det_{q}A$, is defined through the relation
\begin{equation}
{A^{i_{1}}}_{j_{1}} \ldots {A^{i_{N}}}_{j_{N}} \epsilon _{q}^{j_{1} \ldots
j_{N}}= \epsilon _{q}^{i_{1} \ldots i_{N}} det _{q} A, \label{det}
\end{equation}
and this definition, together with (\ref{RAA}), makes $det _{q} A$ commute with
all elements of $A$.

\subsection{The Calculus for $GL_{q}(N)$}
Following the approach of \cite{Z2}, we introduce the exterior derivative $d$
on $GL_{q}(N)$ as a left action which maps $k$-forms to $(k+1)$-forms and
satisfies the same properties as the undeformed exterior derivative, i.e. it is
a linear operator, and for any forms $f$ and $g$,
\begin{eqnarray}
d^{2}g=0, & d(fg)=(df)g+(-1)^{k}f(dg),
\end{eqnarray}
where $f$ is a $k$-form.  Functions of the elements of $A$ are taken as
0-forms,
and we take the elements of $dA$ to be a basis for 1-forms; the commutation
relations (first found in \cite{M2,M3,Ma1,Ma2} but put in R-matrix notation in
\cite{S,S1,S2}) are
\begin{eqnarray}
dA_{1}A_{2}=R_{12}^{-1}A_{2}dA_{1}R_{21}^{-1}, & dA_{1}dA_{2}+R_{12}^{-1}dA_{2}
dA_{1}R_{21}^{-1}=0. \label{diff}
\end{eqnarray}
These are consistent with (\ref{char}) and (\ref{RAA}), of course.
(Alternatively, we could have taken
\begin{equation}
A_{1}dA_{2}=R_{12}^{-1}dA_{2}A_{1}R_{21}^{-1}
\end{equation}
which is also consistent with (\ref{char}) and (\ref{RAA}), and gives the same
$dA-dA$ commutation relations as above.)

{\bf Aside:} It is convenient to introduce the numerical matrix $D$ given by
\begin{equation}
D\equiv q^{2N-1}tr_{2}(P_{12}\tilde{R}_{12})=diag(1,q^{2},\ldots ,q^{2(N-1)
}),
\end{equation}
where $\tilde{K}=[(K^{t_{1}})^{-1}]^{t_{1}}$ for any $N^{2} \times N^{2}$
matrix $K$.  (Here $tr_{I}$ and $t_{I}$ denote tracing and transposing
with respect to the $I$th pair of indices, respectively.)  The definition of
the $D$-matrix, together with (\ref{RAA}), gives
\begin{equation}
(D^{-1})^{t}A^{t}D^{t}S(A)^{t}=S(A)^{t}(D^{-1})^{t}A^{t}D^{t}=1,
\label{DADA}
\end{equation}
where $S(A)$, the antipode of $A$, is simply $A^{-1}$.  A consequence of
this relation is that
\begin{equation}
S^{2}(A)=DAD^{-1}.
\end{equation}
(\ref{RAA}) and (\ref{DADA}) together imply the identities
\begin{eqnarray}
\tilde{R}_{12}=D_{1}^{-1}R_{12}^{-1}D_{1}, & \tilde{R}_{21}=D_{1} R_{21}^{-1}
D^{-1}_{1}.
\end{eqnarray}
All of the above implies two important results:  if $M$ is an $N \times N$
matrix, then
\begin{eqnarray}
tr_{1}(D_{1}^{-1}R_{12}^{-1}M_{1}R_{12})^{i}{}_{j}&=&tr_{1}(D_{1}^{-1}R_{21}
M_{1}R_{21}^{-1})^{i}{}_{j}\nonumber \\
&=& tr(D^{-1}M) \delta ^{i}_{j},
\end{eqnarray}
and if the elements of $M$ commute with the elements of $A$,
\begin{equation}
tr(D^{-1}S(A)MA)=tr(D^{-1}M).
\end{equation}
For this reason, $tr(D^{-1}M)$ is called the invariant trace of $M$.

(\ref{RAA}) and (\ref{diff}) go into themselves under the right coaction $A
\mapsto A A^{\prime}$ and the left coaction $A \mapsto A^{\prime} A$, where
$A^{\prime}$ is a constant (i.e. $dA^{\prime}$=0) $GL_{q}(N)$ matrix satisfying
(\ref{RAA}), and whose elements commute with those of $A$ and $dA$.  $d$ is
invariant under both these coactions.  However, the Cartan-Maurer form
\begin{equation}
\Omega \equiv S(A) dA
\end{equation}
is left-invariant and right-covariant i.e. $\Omega \mapsto \Omega$ and $\Omega
\mapsto S(A^{\prime}) \Omega A^{\prime}$ under the respective coactions above.
(We could have chosen the left-covariant, right-invariant form $dA\,S(A)$
instead.)  This allows us to define the left- and right-invariant 1-form $\xi$
as
\begin{equation}
\xi \equiv -q^{2N-1}tr(D^{-1} \Omega).
\end{equation}
$\Omega$ satisfies the following equations due to (\ref{RAA}) and
(\ref{diff}):
\begin{eqnarray}
\Omega _{1} A_{2} &=& A_{2} R_{12}^{-1} \Omega _{1} R_{21}^{-1},\nonumber \\
\Omega _{1}dA_{2}+dA_{2} R_{12}^{-1} \Omega _{1} R_{12}&=&0, \nonumber \\
\Omega _{1} R_{21}^{-1} \Omega _{2} R_{21}+R_{21}^{-1} \Omega _{2} R_{12}^{-1}
\Omega _{1}&=&0.\label{omega}
\end{eqnarray}
Using these, (\ref{char}), and the definition of $D$,
\begin{eqnarray}
dA = \lambda^{-1}[\xi, A], & d\Omega=-\Omega^{2}=\lambda^{-1}\{\xi,
\Omega\},
\end{eqnarray}
so $\xi$ is in fact the generator of the exterior derivative.  These imply
that the exterior derivative of any form $f$ is given by
\begin{equation}
df=\lambda^{-1}[\xi,f]_{\pm}
\end{equation}
(where $[\, ,\, ]_{\pm}$ is a commutator for even-forms, an anticommutator
for odd-forms).  $det_{q}A$ is a 0-form,  and the above equations imply that
\begin{eqnarray}
\Omega (det_{q}A) = q^{-2} (det_{q}A) \Omega, \nonumber \\
d(det_{q}A)=-q^{-1}(det_{q}A) \xi=-q \xi (det_{q}A).
\end{eqnarray}
(A consequence of these equations is that both $d\xi$ and $\xi^{2}$ vanish.)
The elements of $\Omega$ form a linearly independent basis for 1-forms, and we
shall use them instead of the elements of $dA$ from now on.

We now introduce the inner derivation, which we take to be a left action
mapping
$k$-forms to $(k-1)$-forms.  Its action on the $N^{2}$ elements of $A$ and
$\Omega$ is given by introducing $N^{2}$ vector fields $X^{i}{}_{j}$, and the
associated $N^{2}$ inner derivations are the entries in the matrix $i_{X}$
whose
elements are
\begin{equation}
(i_{X})^{i}{}_{j}=i_{X^{i}{}_{j}}.
\end{equation}
$i_{X}$ must act on 0- and 1-forms in a way preserving the commutation
relations
(\ref{RAA}) and (\ref{omega}); the appropriate actions are
\begin{eqnarray}
i_{X_{1}}A_{2}&=&A_{2}R_{21}i_{X_{1}}R_{12},\nonumber \\
R_{21}i_{X_{1}}R_{12}\Omega _{2} +\Omega_{2} R_{21}i_{X_{1}}R_{12} &=&\frac{1-
R_{21}R_{12}}{\lambda}.\label{inner}
\end{eqnarray}
This last relation implies
\begin{equation}
i_{X} \xi +\xi i_{X}=I.
\end{equation}
(Notice that by using (\ref{char}), $\frac{1-R_{21}R_{12}}{\lambda}$ could be
replaced by $-\hat{R}_{12}$ if so desired.)  On $det_{q}A$, the inner
derivation acts as
\begin{equation}
i_{X}(det_{q}A)=q^{2}(det_{q}A)i_{X}.
\end{equation}
The commutation relations between the inner derivation matrices are similar
to the ones for $\Omega$:
\begin{equation}
R_{12}^{-1}i_{X_{1}}R_{12}\, i_{X_{2}}+i_{X_{2}}R_{21}i_{X_{1}}R_{12}=0.
\label{RiRi}
\end{equation}
Equations (\ref{inner}) imply that $i_{X}$ is left-invariant and
right-covariant
under the respective coactions on $A$.

We may now introduce the Lie derivative matrix $L_{X}$ in the same way as in
the classical theory, i.e. a left action taking $k$-forms to $k$-forms given by
\begin{equation}
L_{X}\equiv i_{X}d+di_{X},
\end{equation}
where $L_{X}$ is a matrix with elements $L_{X^{i}{}_{j}}$ which by definition
transforms in the same way as $i_{X}$ does.  The equations already given for
$d$ and $i_{X}$ imply a whole host of relations involving $L_{X}$:
\begin{eqnarray}
L_{X}d &=&  dL_{X},\nonumber \\
R_{21}L_{X_{1}}R_{12}\, i_{X_{2}}-i_{X_{2}}R_{21}L_{X_{1}}R_{12}&=&\lambda^{-
1}(R_{21}R_{12}i_{X_{2}}-i_{X_{2}}R_{21}R_{12}), \nonumber \\
R_{21}L_{X_{1}}R_{12}L_{X_{2}} - L_{X_{2}}R_{21}L_{X_{1}}R_{12}&=&\lambda^{-
1}(R_{21}R_{12}L_{X_{2}} -L_{X_{2}}R_{21}R_{12}), \nonumber \\
L_{X_{1}}A_{2}&=&A_{2}R_{21}L_{X_{1}}R_{12}+A_{2}(\frac{1-R_{21}R_{12}}
{\lambda}),\nonumber \\
R_{21}L_{X_{1}}R_{12}\Omega_{2}-\Omega_{2}R_{21}L_{X_{1}}R_{12}&=&
\lambda^{-1}(R_{21}R_{12}\Omega_{2}-\Omega_{2}R_{21}R_{12}),\nonumber \\
L_{X}\xi&=&\xi L_{X},\label{Lie}
\end{eqnarray}
and for the determinant,
\begin{equation}
L_{X}(det_{q}A)=q^{2}(det_{q}A)L_{X}-q(det_{q}A).
\end{equation}
Many of these relations take a much simpler form if we introduce the
Lie derivative valued operator $Y$ given by
\begin{equation}
Y=1-\lambda L_{X},
\end{equation}
which, of course, has the same transformation properties as $L_{X}$.
Using this, we obtain
\begin{eqnarray}
Yd&=&dY,\nonumber \\
R_{21}Y_{1}R_{12}\, i_{X_{2}}&=&i_{X_{2}}R_{21}Y_{1}R_{12},\nonumber \\
R_{21}Y_{1}R_{12}Y_{2}&=&Y_{2}R_{21}Y_{1}R_{12},\nonumber \\
Y_{1}A_{2}&=&A_{2}R_{21}Y_{1}R_{12},\nonumber \\
R_{21}Y_{1}R_{12}\Omega_{2}&=&\Omega_{2}R_{21}Y_{1}R_{12},\nonumber \\
Y\xi&=&\xi Y,\label{Y}
\end{eqnarray}
and
\begin{equation}
Y(det_{q}A)=q^{2}(det_{q}A)Y.
\end{equation}
$Y$ is useful for more than making pretty equations.  Since its leading term
is unity, it is invertible.  More importantly, we can define a quantity
$Det\, Y$, which we identify as the determinant of $Y$, satisfying
\begin{equation}
Y(Det\, Y)=(Det\, Y)Y.
\end{equation}
This quantity is defined through \cite{SWZ}
\begin{equation}
(Y_{1 \ldots N}^{(1)} \ldots Y_{1 \ldots N}^{(N)})^{i_{1} \ldots
i_{N}}{}_{j_{1} \ldots j_{N}} \epsilon_{q} ^{j_{1} \ldots j_{N}}=
\epsilon_{q}^{i_{1} \ldots i_{N}}Det \, Y , \label{Dety}
\end{equation}
where
\begin{equation}
Y^{(k)}_{1 \ldots N}=\left \{
	\begin{array}{ll}
	(R_{kN} \ldots R_{k(k+1)})^{-1}Y_{k}(R_{kN} \ldots R_{k(k+1)}) &
	\mbox{for $k=1,\ldots ,N-1$,} \\
	Y_{N} & \mbox{for $k=N$.}
	\end{array}
	\right.
\end{equation}
This determinant is invariant under transformations of $Y$  (i.e. $Y \mapsto
Y$ for $A \mapsto AA^{\prime}$ and $Y \mapsto S(A^{\prime})YA^{\prime}$ for
$A \mapsto A^{\prime}A$, with $Y$ and $A^{\prime}$ having commuting elements),
and satisfies the following as a consequence of the above equations:
\begin{eqnarray}
d (Det\, Y)&=&(Det\, Y)d,\nonumber \\
(Det\, Y)i_{X}&=&i_{X}(Det\, Y),\nonumber \\
(Det\, Y)A&=&q^{2}A(Det\, Y),\nonumber \\
(Det\, Y)\Omega &=& \Omega (Det\, Y),\nonumber \\
(Det\, Y)\xi &=&\xi (Det\, Y),\nonumber \\
(Det\, Y)(det_{q}A)&=&q^{2N}(det_{q}A)(Det\, Y).
\end{eqnarray}
The above equations for $Det\, Y$ suggest the definition of an operator
$H_{0}$ as
\begin{equation}
Det \, Y
 \equiv q^{2H_{0}}=1+q\lambda [H_{0}]_{q}
\end{equation}
where
\begin{equation}
[x]_{q}=\frac{1-q^{2x}}{1-q^{2}}.\label{bra}
\end{equation}
$H_{0}$ commutes with $Y$, $d$, $i_{X}$, $\Omega$, and $\xi$, and satisfies
\begin{eqnarray}
[H_{0},A]=A, & [H_{0},det_{q}A]=N(det_{q}A).\label{Hcomm}
\end{eqnarray}
This operator will be important in the next section.

\section{$SL_{q}(N)$}
\subsection{The Quantum Group $SL_{q}(N)$}

There seems to be an obvious way to specify the quantum group $SL_{q}(N)$:
take the matrix $A$ and set its determinant to unity.  Unfortunately, this
doesn't work.  True, $det_{q}A$ as defined in the previous section commutes
with the elements of $A$, but it does {\em not} commute with such quantities
as $\Omega$ and $Y$.  Therefore, instead of imposing $det_{q}A=1$, we {\em
define} matrices $T$ as
\begin{equation}
T=(det_{q}A)^{-\frac{1}{N}}A.\label{defT}
\end{equation}
With $det_{q}T$ defined as in (\ref{det}), the centrality of $det_{q}A$
automatically gives $T$ determinant unity.  Furthermore, the antipode of $T$
is also given by $T^{-1}$.  Therefore, this matrix $T$ is what we
identify as an element of the defining representation of $SL_{q}(N)$, since
it also satisfies (\ref{RAA}) with $A$ replaced by $T$.  However, as we will
see in the next section, it becomes convenient to introduce the matrix
\begin{equation}
{\cal R}_{12} = q^{-\frac{1}{N}}R_{12},
\end{equation}
which we identify as the R-matrix for $SL_{q}(N)$.  Thus, we shall write
(\ref{RAA}) as
\begin{equation}
{\cal R}_{12}T_{1}T_{2}=T_{2}T_{1}{\cal R}_{12}.
\end{equation}

\subsection{The Calculus for $SL_{q}(N)$}
The exterior derivative on $SL_{q}(N)$ can be taken to be the same as that
introduced on $GL_{q}(N)$; this is because $T$ is a function of elements
of $A$, so its differentials are given by
\begin{equation}
dT=\lambda^{-1}[\xi ,T].
\end{equation}
Note that this implies that the Cartan-Maurer form $\tilde{\Omega}$ for
$SL_{q}(N)$ is given by
\begin{equation}
\tilde{\Omega}\equiv S(T)dT=q^{\frac{2}{N}}\Omega +q \, [1/N]_{q}\xi.
\footnote{This relation implies that the matrix of differential forms
introduced in \cite{Z2} is equal to $-q^{2N-1}\Omega$.}
\label{SLform}
\end{equation}
(see (\ref{bra}) for the definition of $[\, ]_{q}$.)
In the classical limit $q \rightarrow 1$, $\tilde{\Omega}$ is
traceless, giving the appropriate reduction from $N^{2}$ to $N^{2}-1$
independent elements in the Cartan-Maurer matrix 1-form for $SL(N)$.

We have thus found a way to set the determinant of our $SL_{q}(N)$ matrices to
unity; for the calculus of the group, we must do something similar, namely
impose a constraint so that the number of independent differential operators is
reduced from $N^{2}$ to $N^{2}-1$.  In a way, we have already done this,
because
(\ref{Hcomm}) and (\ref{defT}) together imply
\begin{equation}
[H_{0},T]=0,
\end{equation}
so that $H_{0}$ commutes with everything of interest in $SL_{q}(N)$, i.e.
matrices, forms, exterior derivative, etc.  Thus, within the context of
$SL_{q}(N)$, $H_{0}$ is irrelevant, reducing the number of generators
from $N^{2}$ to $N^{2}-1$, as desired.  Explicitly, this restriction is
accomplished by defining a new Lie derivative valued operator $Z$ by
\begin{equation}
Z\equiv q^{-\frac{2H_{0}}{N}}Y,\footnote{When restricted to acting on 0-forms,
this operator is identical to the operator $Y$ in \cite{Z2}.}
\end{equation}
Note that the determinant of $Z$, computed using (\ref{Dety}), is unity.  This
is equivalent to the introduction of a set of $N^{2}$ ``vector fields''
$V^{i}{}_{j}$ through $Z=1-\lambda L_{V}$, so that
\begin{equation}
L_{V}=L_{X}+q^{-1}[H_{0}/N]_{q^{-1}}-q^{-1}\lambda L_{X}[H_{0}/N]_{q^{-1}}.
\end{equation}
The fact that $Det\, Z=1$ implies that only $N^{2}-1$ of the elements of
$L_{V}$
are actually independent, which is precisely what we require for $SL_{q}(N)$.
In the classical limit, $H_{0}=-tr(L_{X})$,  so $L_{V}$ becomes traceless;
thus,
$V$ contains only $N^{2}-1$ linearly independent vector fields, as we'd expect.

Now that we have obtained all these quantities, we want to find the various
relations they satisfy.  As a starting point, note that
the commutation relations between $\Omega$ and $T$ are given by
\begin{equation}
\Omega_{1}T_{2}=q^{\frac{2}{N}}T_{2}R_{12}^{-1}\Omega_{1}R_{21}^{-1}
=T_{2}{\cal R}_{12}^{-1}\Omega_{1}{\cal R}_{21}^{-1}.
\end{equation}
Here we see the appearance of ${\cal R}_{12}$, as promised.  In fact, there
is a general pattern:  by using the substitutions $A \rightarrow T$, $R_{12}
\rightarrow {\cal R}_{12}$, and $L_{X} \rightarrow L_{V}$, we obtain
most of the corresponding relations for $SL_{q}(N)$.  ($\Omega$ remains
unchanged, so the last of equations (\ref{omega}) does not have ${\cal R}
_{12}$ in place of $R_{12}$.)  $L_{V}$ satisfies
\begin{eqnarray}
{\cal R}_{21}L_{V_{1}}{\cal R}_{12}L_{V_{2}}-L_{V_{2}}{\cal
R}_{21}L_{V_{1}}{\cal R}_{12}&=&\lambda^{-1}({\cal R}_{21}{\cal
R}_{12}L_{V_{2}}-L_{V_{2}}{\cal R}_{21}{\cal R}_{12}),\nonumber \\
{\cal R}_{21}L_{V_{1}}{\cal R}_{12}i_{X_{2}}-i_{X_{2}}{\cal
R}_{21}L_{V_{1}}{\cal R}_{12}&=&\lambda^{-1}({\cal R}_{21}{\cal
R}_{12}i_{X_{2}}-i_{X_{2}}{\cal R}_{21}{\cal R}_{12}).
\end{eqnarray}
The actions of the various operators on the 0- and 1-forms of $SL_{q}(N)$
are given by
\begin{eqnarray}
L_{V_{1}}T_{2}&=&T_{2}{\cal R}_{21}L_{V_{1}}{\cal
R}_{12}+T_{2}(\frac{1-{\cal R}_{21}{\cal R}_{12}}{\lambda}), \nonumber \\
{\cal R}_{21}L_{V_{1}}{\cal R}_{12}\Omega _{2}-\Omega_{2}{\cal R}_{21}
L_{V_{1}}{\cal R}_{12}&=&\lambda^{-1}({\cal R}_{21}{\cal R}_{12}\Omega_{2}
-\Omega_{2}{\cal R}_{21}{\cal R}_{12}), \nonumber \\
i_{X_{1}}T_{2}&=&T_{2}{\cal R}_{21}i_{X_{1}}{\cal R}_{12}, \nonumber \\
{\cal R}_{21}i_{X_{1}}{\cal R}_{12}\tilde{\Omega}_{2}+\tilde{\Omega}_{2}
{\cal R}_{21}i_{X_{1}}{\cal R}_{12}&=&\frac{1-{\cal R}_{21}{\cal R}_{12}}
{\lambda}.
\end{eqnarray}
As a consequence, $\xi$ satisfies
\begin{equation}
L_{V}\xi = \xi L_{V}.
\end{equation}
The relations for $Z$ corresponding to (\ref{Y}) are easily obtained by using
$L_{V}=\frac{1-Z}{\lambda}$ in all of the above equations.

\section{Conclusion}
Many of the relations in this work are not unique to a discussion of
$GL_{q}(N)$
or $SL_{q}(N)$; for instance, given an R-matrix, the differentials of the
quantum hyperplane can be defined and their commutation relations found, and
the
corresponding Levi-Civita tensor found.  This allows the definition of the
determinant of a quantum matrix.  However, much of this paper is based on the
fact that the characteristic equations for $GL_{q}(N)$ and $SL_{q}(N)$ are
quadratic in $\hat{R}_{12}$.  For other quantum groups, the characteristic
equation for $\hat{R}_{12}$ may be of higher degree (as in $SO_{q}(N)$ and
$Sp_{q}(N)$, where the characteristic equations are cubic), in which case
relations like (\ref{diff}) will not be consistent with (\ref{RAA}).  Although
others have looked at the calculi of such quantum groups \cite{CW}, it remains
to be seen whether it is possible to use techniques similar to ours in these
cases.

\end{document}